\documentclass[prd,preprintnumbers,aps,showpacs,tightenlines,notitlepage,nofootinbib,superscriptaddress]{revtex4-1}

\usepackage{graphicx}
\usepackage{amssymb}
\usepackage{amsmath}
\usepackage{bm}
\usepackage{slashed}
\usepackage{datetime}
\usepackage{epsfig}

 \newcommand{\be}{\begin{equation}}
\newcommand{\ee}{\end{equation}}
\newcommand{\ba}{\begin{eqnarray}}
\newcommand{\ea}{\end{eqnarray}}
\newcommand{\la}{\langle}
\newcommand{\ra}{\rangle}

 \newcommand{\cdott}{{\mskip -1.5mu} \cdot {\mskip -1.5mu}}
 \def\C{_{_C}}
 
\begin{document}


\preprint{JLAB-THY-14-1885}


\title{Extraction of the  distribution function $h^{\perp}_{1T}$ from experimental data}

\author{Christopher Lefky}
\email{ChristopherLefky@creighton.edu}
\affiliation{Creighton University, Omaha, Nebraska 68102, USA}

\author{Alexei Prokudin}
\email{prokudin@jlab.org}
\affiliation{Jefferson Lab, 12000 Jefferson Avenue, Newport News, Virginia
  23606, USA}

\date{\today}
 

\begin{abstract}
\noindent
   We attempt an extraction of the pretzelosity distribution ($h^{\perp}_{1T}$) from preliminary COMPASS, HERMES, and JLAB experimental data on
$\sin(3\phi_h - \phi_S)$ asymmetry on proton, and effective  deuteron and neutron targets.  
The resulting distributions, albeit with big errors, for the first time 
show tendency for up-quark pretzelosity to be positive and down-quark pretzelosity to be negative. 
A model relation of
pretzelosity distribution and orbital angular momentum of quarks is used to estimate contributions of 
up and down quarks. 
\end{abstract}

\pacs{13.88.+e, 
      13.85.Ni, 
      13.60.-r, 
      13.85.Qk} 
\keywords{Semi-inclusive deep inelastic scattering, 
	  transverse momentum dependent distribution functions, 
	  spin asymmetries}
	  
\maketitle


\section{Introduction}
\label{Sec-1:introduction}
\noindent
The proton is a very intricate dynamical system of quarks and gluons. Spin decomposition and partonic structure of the nucleon remain key problems of modern nuclear physics and
 orbital angular momentum (OAM) of partons has 
emerged as an essential part of our understanding of the internal structure of the nucleon. Studying the structure of the proton is one of the main goals 
of many past and present experimental facilities and experiments such as H1 (DESY),  ZEUS (DESY), HERMES (DESY), COMPASS (CERN), Jefferson Lab, RHIC (BNL), 
various Drell-Yan experiments ~\cite{Peng:2014hta}, and $e^+e^-$ annihilation experiments by the Belle and {\it BABAR} Collaborations. Future Jefferson Lab 12 \cite{Dudek:2012vr} and EIC \cite{Accardi:2012qut} studies are going to provide very detailed experimental data
that will improve our knowledge of hadron structure in valence and sea regions.
Description of semi-inlcusive deep inelastic scattering (SIDIS),  $e^+e^-$ annihilation to two hadrons, and Drell-Yan process at low transverse momentum (with respect to resolution scale) of 
observed particles is achieved in terms of so-called transverse momentum dependent distribution and fragmentation functions (collectively called TMDs). TMDs depend on the longitudinal momentum fraction and on the transverse motion vector of partons inside of the nucleon and thus 
allow for three-dimensional ``3-D" representation of the nucleon structure in momentum space and are related to the OAM of partons.

One particular 
TMD distribution function might play a role in our understanding of the spin of the nucleon.
This distribution is called pretzelosity ($h^{\perp}_{1T}$), and its name stems from the fact that a  polarized proton might not be
spherically symmetric \cite{Miller:2003sa}. This function depends on the fraction of hadron momentum carried by the parton, $x$, and the intrinsic transverse momentum of the parton, $\bm{k}_\perp$, and  corresponds to a quadrupole modulation of parton density in the distribution of transversely polarized quarks
in a transversely polarized nucleon \cite{Mulders:1995dh,Goeke:2005hb,Bacchetta:2006tn}
\begin{align}  
 \Phi^{[i \sigma^{\alpha +}\gamma_5]}(x,\bm{k}_\perp)  = 
 S_{T}^{\alpha} \, h_{1}
(x,{k}_{\perp}^{2}) 
 + S_L\,\frac{k_{\perp}^{\alpha}}{M} \, h_{1L}^{\perp}
(x,{k}_{\perp}^{2}) 
 - \frac{k_{\perp}^{\alpha} k_{\perp}^{\rho}
     -\frac{1}{2}\,{k}_\perp^{2}\,g_T^{\alpha\rho}}{M^2}\, S_{T \rho} 
   h_{1T}^{\perp}
(x,{k}_{\perp}^{2}) 
- \frac{\epsilon_{T}^{\alpha\rho} k_{\perp \rho}^{}}{M}  
   h_{1}^{\perp} 
(x,{k}_{\perp}^{2}) \, .
\label{eq:phi}
 \end{align}   
In this formula $S_T$ and $S_L$ are transverse and longitudinal components of polarsation vector and other functions that enter in the projection of parton density with $\sigma^{\alpha +}\gamma_5$ are transversity \cite{Ralston:1979ys} ($h_{1}$), Boer-Mulders function \cite{Boer:1997nt} ($h_{1}^{\perp}$), and so-called worm-gear or Kotzinian-Mulders function  \cite{Kotzinian:1997wt} ($h_{1L}^{\perp}$). As one can see, the pretzelosity distribution enters with $k_{\perp}^{\alpha} k_{\perp}^{\rho}
     -\frac{1}{2}\,{k}_\perp^{2}\,g_T^{\alpha\rho}$ coefficient  that corresponds to a quadrupole modulation of parton density in momentum space.
The pretzelosity distribution in convolution with the Collins fragmentation function \cite{Collins:1992kk} generates $\sin(3\phi_h - \phi_S)$ asymmetry in Semi-Inclusive Deep Inelastic Scattering (SIDIS) and was studied experimentally by COMPASS \cite{Parsamyan:2007ju,Parsamyan:2010se,Parsamyan:2013ug,Parsamyan:2013fia}
and HERMES \cite{Diefenthaler:2010zz,Schnell:2010zza,Pappalardo:2010zz} collaborations and JLAB  \cite{Zhang:2013dow}. We attempt the first extraction of pretzelosity from the latest experimental data
 \cite{Parsamyan:2007ju,Parsamyan:2010se,Parsamyan:2013ug,Parsamyan:2013fia,Diefenthaler:2010zz,Schnell:2010zza,Pappalardo:2010zz,Zhang:2013dow} using extracted Collins fragmentation function from Ref.~\cite{Anselmino:2013vqa} for our analysis. we are going to use tree level approximation
and neglect possible effects of TMD evolution (such as Sudakov suppression) in this paper and include only the relevant DGLAP  (Dokshitzer-Gribov-Lipatov-Altarelli-Parisi) evolution of collinear quantities. In fact, the span of $Q^2$ in the experimental data is narrow enough to assume small possible effects from evolution.

Model calculations of pretzelosity including the bag and Light-Cone Quark
 models and predictions for experiments are presented in Refs.~\cite{Kotzinian:2008fe,She:2009jq,Avakian:2008dz,Pasquini:2008ax,Pasquini:2009bv,Boffi:2009sh,Avakian:2010br}. Note that most  models predict negative $u$-quark and positive $d$-quark pretzelosity.
  
 In a vast class of models with a spherically symmetric nucleon wave function in the rest frame, the pretzelosity distribution is related to the OAM of quarks
by the following relation \cite{She:2009jq,Avakian:2010br,Efremov:2010cy}
\begin{align}
{\cal L}_{z}^a = -\int d x \,  d^2 \bm{k}_{\perp} \, \frac{k_\perp^2}{2M^2} h_{1T}^{\perp a}(x, k_\perp 
^2) = -\int d x  \; h_{1T}^{\perp (1) a}(x) \;. 
\label{oam}
\end{align}

It was shown in Ref.~\cite{Lorce:2011kn} that the relation of Eq.~\eqref{oam} did not correspond to the intrinsic OAM of quarks. This relation is valid on the amplitude level and not on the operator level and may hold only numerically~\cite{Lorce:2011kn}
as OAM is chiral and charge even, but pretzelosity is chiral and charge odd. 
We warn the reader that the relation of Eq.~\eqref{oam} is model dependent and thus one cannot derive solid conclusions based on it; nevertheless, it appears very interesting to attempt an extraction of the pretzelosity distribution ($h^{\perp}_{1T}$) from the experimental data  on
$\sin(3\phi_h - \phi_S)$ asymmetry and compare numerical results of Eq.~\eqref{oam} with existing calculations of OAM.

Lattice QCD
results \cite{Hagler:2007xi,Syritsyn:2011vk,Liu:2012nz}  use Ji's relation of Ref.~\cite{Ji:1996ek} of the total angular momentum of the flavor $q$ contribution to the spin of the nucleon and GPDs $H_q$ and $E_q$:
 $J_q = \frac{1}{2}\int dx x [ H_q(x,0,0) + E_q(x,0,0)]$. Contribution of the quark spin is then subtracted from the result: $L_z^q = J_q - \frac{1}{2} \Sigma_q$. References~\cite{Hagler:2007xi,Syritsyn:2011vk,Liu:2012nz} use only the so-called connected insertions in lattice simulations and find the following result:
\begin{align}
 L_{z}^u <0, \;L_{z}^d >0, \;  |L_{z}^{u+d}| \ll |L_{z}^{u}|, |L_{z}^{d}| \, . 
\label{lattice}
\end{align}
 
Reference~\cite{Deka:2013zha} shows that, while $|L_{z}^{u+d}|$ largely cancels if only connected insertions are considered, the results change when disconnected insertions  are included: their contributions  are large and positive. Results of Ref.~\cite{Deka:2013zha} at $Q^2=4$ GeV$^2$ imply that
\begin{align}
 L_{z}^u <0, \;L_{z}^d >0 \, , 
\label{lattice1}
\end{align}
thus the smallness of the total $u+d$ contribution to the OAM is not confirmed. The connected insertions do not affect either the difference $L_{z}^u-L_{z}^d$ or the sign of $L_{z}^u$, $L_{z}^d$.

It is worth mentioning that TMDs exhibit the so-called generalized universality. Some TMDs may depend on the process. The most notorious examples are Sivers \cite{Sivers:1989cc,Boer:1997nt} and Boer-Mulders
\cite{Boer:1997nt} distributions   that have opposite signs in SIDIS and DY \cite{Brodsky:2002rv,Collins:2002kn,Kang:2009bp}. Apart from the sign, these functions are the same and universal;  however, it turns out that  there might be several universal functions corresponding to pretzelosity distribution. In particular, it was found in Ref.~\cite{Buffing:2012sz} that
there are three different universal functions corresponding to pretzelosity. Those functions are in principle accessible in various processes, but one cannot distinguish among them in SIDIS;
thus, we will use only one function $h^{\perp}_{1T} \equiv {h^{\perp}_{1T}}^{\rm SIDIS}$.

As we mentioned previously, the relation of Eq.~\eqref{oam} OAM and the pretzelosity function is a model inspired relation. It was shown that
OAM is related to the so-called generalised transverse momentum dependent distributions (GTMDs), in particular to one denoted as $F_{1,4}$ \cite{Meissner:2009ww}. There are two ways of constructing the OAM of quarks, depending on the configuration of the gauge link in the operator definition: either the canonical OAM of Jaffe-Manohar \cite{Jaffe:1989jz} in the spin decomposition $\frac{1}{2} = \frac{1}{2} \Delta \Sigma + L_z^q + \Delta G + L_z^g$ or
the kinetic OAM in the definition of Ji \cite{Ji:1996ek} in the spin decomposition  $\frac{1}{2} = \frac{1}{2} \Delta \Sigma + L_z^q + J^g$.  The definition of 
OAM in these two decompositions differs by the presence of the derivative $i \partial$ in the definition of Jaffe-Manohar \cite{Jaffe:1989jz} and the covariant derivative $i D = i \partial - g A$
in the definition of Ji \cite{Ji:1996ek}. The presence of gauge field in kinetic OAM makes it different from canonical OAM.
The relation of $F_{1,4}$ and OAM of partons in a longitudinally polarized nucleon was shown in Ref.~\cite{Lorce:2011kd} and model and QCD calculations of the canonical and the kinetic OAM were performed in Refs.~\cite{BC:2011dv,Kanazawa:2014nha,Mukherjee:2014nya}.
Results of Ref.~\cite{Mukherjee:2014nya} indicate that the total kinetic OAM in a one-quark model $L_z^q \sim (-0.04 \div -0.1)$ at $Q=3$ GeV. Canonical and kinetic OAM were studied in Refs.~\cite{Courtoy:2013oaa} and the model results suggest that for a $u$ quark canonical $L_z^u = 0.11$ and kinetic $L_z^u = 0.13$ at the hadronic scale.
 The numerical difference between two definitions is generated by the presence of the gauge field: in models without the gauge field term, such as the scalar diquark model, one obtains the same value for kinetic and canonical OAM  \cite{Burkardt:2008ua}. Model results of Refs.~\cite{Burkardt:2008ua,Courtoy:2013oaa} are 
of the opposite sign of lattice QCD results that suggest $L_z^u<0$ and $L_z^d>0$.

The rest of the paper is organized as follows: in Section~\ref{sectionI} we will derive a general formula for the $A_{UT}^{\sin(3\phi_h-\phi_S)}$ single spin asymmetry associated with pretzelosity in TMD formalism. Formulas for unpolarized and polarized cross sections will be presented in Sections \ref{sectionIIA} and
 \ref{sectionIIB}. We will calculate the probability that existing experimental data from COMPASS, HERMES, and JLAB indicate that all pretzelosity functions are exactly equal to {\it zero}, i.e. the so-called null-signal hypothesis, in Section~\ref{sectionIII}. Then we attempt a detailed phenomenological fit of pretzelosity distributions in Section~\ref{sectionIV} where we present resulting parameters of the fit of pretzelosity distributions and comparison with existing data. We will give predictions for future measurements at Jefferson Lab 12 in Section~\ref{sectionIVA}. We will compare resulting pretzelosity distribution to models in Section~\ref{sectionV} and test model relations on pretzelosity in Section~\ref{sectionVI}. Using the model relation of the OAM of quarks and pretzelosity we will calculate the OAM of up and down quarks in Section~\ref{sectionVII}. Finally we will conclude in Section~\ref{conclusions}. 
\section{$A_{UT}^{\sin(3\phi_h-\phi_S)}$ Single Spin Asymmetry\label{sectionI}}
\noindent
 The part of the SIDIS cross section  we are interested in reads~\cite{Mulders:1995dh,Bacchetta:2006tn,Anselmino:2011ch}
\begin{align}
\frac{d\sigma}{dx\, dy\, d\phi_S \,dz\, d\phi_h\, d P_{hT}}
= 
\frac{\alpha^2 2 P_{hT}}{x y\, Q^2}\,
\left\{
\left(1-y+\frac{1}{2}y^2\right)\left( F_{UU ,T} + \varepsilon F_{UU ,L}\right)  +  S_T \, (1-y) \, \sin(3\phi_h-\phi_S)\,
F_{UT}^{\sin(3\phi_h -\phi_S)}  + ...\right\}
\label{eq:asymmetry}
\end{align}
where one uses the following standard variables
\be
x = \frac{Q^2}{2\,P\cdott q}, \;\;
y = \frac{P \cdott q}{P \cdott l}, \;\;
z = \frac{P \cdott P_h}{P\cdott q}, \; \varepsilon \approx \frac{1-y}{1-y+\frac{1}{2}y^2},
  \label{eq:xyz}
\ee
where $\alpha$ is the fine structure constant, $Q^2 = -q^2 = -(l-l')^2$ is the virtuality of the exchanged photon, $\bm{P}_{h T}$ is the transverse momentum of the produced hadron, $S_T$ is transverse polarization, and $\phi_h,\phi_S$ are the azimuthal angles of the produced hadron and the polarization vector with respect to the lepton scattering plane formed by $\bm{l}$ and $\bm{l}'$.
 $F_{UU ,L}= 0$ at ${\cal O} (k_\perp/Q)$ order of accuracy. Structure functions that we are interested in in this study are unpolarized structure function $F_{UU ,T}$ and
 spin structure function $F_{UT}^{\sin(3\phi_h -\phi_S)}$; the polarization state of the beam and target are explicitly denoted in definition of structure functions as``$U$"for unpolarized and ``$T$" for transversely polarized. The ellipsis in Eq.~(\ref{eq:asymmetry}) denotes contributions from other spin structure functions.

In this paper we will use the convention of Refs.~\cite{Anselmino:2006rv,Boer:2011fh,Anselmino:2011ch} for the transverse momentum of an incoming quark with respect to the proton's momentum and
the hadron momentum with respect to the fragmenting quark:
\begin{align}
\bm{k}_\perp, \,
\bm{p}_\perp.
\end{align}
The advantage of this convention is that the fragmentation function has a probabilistic interpretation with respect to  vector $\bm{p}_\perp$, i.e.,
\begin{align}
D_{h/a}(z) \equiv  \int d^2 \bm{p}_{\perp} D_{h/a}(z, p_\perp^2)\;.
\end{align}

The structure functions involved in Eq.~\eqref{eq:asymmetry} are convolutions of the distribution and fragmentation functions
$f$ and $D$  \cite{Mulders:1995dh,Bacchetta:2006tn}:
\be
F_{AB}
 = {\cal C}[
    w\, f\, D \big],
\ee
where $A,B$ indicate the polarization state of the beam and the target $U,L,T$, and ${\cal C}[...]$ is defined as
\begin{align}
{\cal C}\bigl[ w\, f\, D \bigr]
= x\,
\sum_a e_a^2 \int d^2 \bm{k}_\perp\,  d^2 \bm{p}_\perp^{}
\, \delta^{(2)}\bigl(z\bm{k}_\perp + \bm{p}_\perp^{} - \bm{P}_{h T} \bigr)
\,w\left(\bm{k}_\perp,-\frac{\bm{p}_\perp^{}}{z}\right)\,
f^a(x,k_\perp^2)\,D_{h/a}(z,p_\perp^2) ,
\label{eq:convolution_our}
\end{align}
or integrating over $d^2 \bm{p}_\perp^{}$,
\begin{align}
{\cal C}\bigl[ w\, f\, D \bigr]
= x\,
\sum_a e_a^2 \int d^2 \bm{k}_\perp\, 
\, 
\,w\left(\bm{k}_\perp,-\frac{\left(\bm{P}_{h T} - z\bm{k}_\perp\right)}{z}\right)\,
f^a(x,k_\perp^2)\,D_{h/a}\left(z,(\bm{P}_{h T} - z\bm{k}_\perp)^2\right) .
\label{eq:convolution_our_main}
\end{align}
For the sake of generality we use ``$f$" and ``$D$" functions to denote distribution and fragmentation TMD in formulas in this section.

The kinematical functions, $w$, can be found in Refs.~\cite{Mulders:1995dh,Bacchetta:2006tn,Anselmino:2011ch}. So-called moments of TMDs  are defined accordingly as
\ba
f^{(n) a}(x) &=& \int d^2 \bm{k}_{\perp} \left(\frac{k_\perp^2}{2M^2}\right)^n f^a(x, k_\perp^2)\;, \label{pdf_moments_our_main}\\
D^{(n)}_{h/a}(z) &=& \int d^2 \bm{p}_{\perp} \left(\frac{p_\perp^2}{2z^2M_h^2}\right)^n D_{h/a}(z, p_\perp^2)\; .
\label{moments_our_main}
\ea
One also defines the ``half'' moment by
\ba
D^{(1/2)}_{h/a}(z) &=& \int d^2 \bm{p}_{\perp} \frac{|p_\perp|}{2 z M_h} D_{h/a}(z, p_\perp^2)\; .
\label{half_moments_our_main}
\ea
Single spin asymmetry (SSA) measured experimentally is defined as:
\be
A_{UT}^{\sin(3\phi_h -\phi_S)}(x,z,y,P_{hT}) \equiv \langle 2 \sin(3\phi_h -\phi_S) \rangle = 2\frac{\int d \phi_h d \phi_S \sin(3\phi_h -\phi_S) \left(d\sigma^{\uparrow} - d\sigma^{\downarrow}\right)}{\int d \phi_h d \phi_S  \left(d\sigma^{\uparrow} + d\sigma^{\downarrow}\right)}\;,
\label{eq:ssa}
\ee
where $\uparrow (\downarrow)$ denote opposite transverse polarizations of the target nucleon, $U$ stands for the unpolarized lepton beam, and $T$ for the transverse polarization of the target nucleon.
The numerator and denominator of Eq.~\eqref{eq:ssa}  can be written as 
\ba
d\sigma^{\uparrow} - d\sigma^{\downarrow} &=& \frac{\alpha^2 2 P_{hT}}{s x^2 y^2} 2\left(1 - y\right)\, \sin(3\phi_h -\phi_S )\, 
F_{UT}^{\sin(3\phi_h -\phi_S)} \;, \nonumber \\
d\sigma^{\uparrow} + d\sigma^{\downarrow} &=& \frac{\alpha^2 2 P_{hT}}{s x^2 y^2} \left(1+(1-y)^2\right) F_{UU ,T}\; .
\label{eq:num_denom}
\ea
The final expression for $A_{UT}^{ \sin(3\phi_h -\phi_S)}$ asymmetry  reads
\ba
A_{UT}^{\sin(3\phi_h-\phi_S)}(x,z,y,P_{hT}) = \frac{\frac{\alpha^2 2 P_{hT}}{s x^2 y^2}}{\frac{\alpha^2 2 P_{hT}}{s x^2 y^2}}
\cdot \frac{2 \left(1 - y\right)}{ \left(1+(1-y)^2\right)}
\cdot \frac{
F_{UT}^{\sin(3\phi_h-\phi_S)}}{ F_{UU ,T}}\;.
\label{eq:final}
\ea 
Note that  $D_{NN}\equiv {2 \left(1 - y\right)}/{(1+(1-y)^2)}$ is often factored out from the measured asymmetry.

\subsection{Unpolarized structure function, $F_{UU,T}$\label{sectionIIA}}
\noindent
The partonic interpretation of the unpolarized structure function $F_{UU,T}$ is the following \cite{Mulders:1995dh,Bacchetta:2006tn,Anselmino:2011ch}
\be
F_{UU,T}
 = {\cal C}[
    f_1 D]\; ,
\ee
where $f_1$ and $D$ are unpolarized TMD distribution and fragmentation functions. We have
\be
F_{UU,T}
 =
x\,
\sum_{a=q,\bar q} e_a^2 \int d^2 \bm{k}_\perp \, 
 f^{a}_1(x,k_\perp^2)\,D_{h/a}\left(z,(\bm{P}_{h T} - z\bm{k}_\perp)^2\right).
\label{eq:fuu}
\ee
Following Refs.~\cite{Anselmino:2008jk, Anselmino:2007fs, Anselmino:2013vqa} we assume Gaussian form for $f_{1}^{a}(x,k_{\perp}^2)$ and $D_{1 h/a}(z,p_{\perp}^2)$:
\ba
f^{a}_1(x,k_{\perp}^2) = f^{a}_1(x)\,\frac{1}{\pi \la k_\perp^2 \ra}\,\exp\left(-\frac{\bm{k}_{\perp}^2}{\la k_\perp^2 \ra}\right)\;, \nonumber \\
D_{h/a}(z,p_{\perp}^2) = D_{h/a}(z)\,\frac{1}{\pi \la p_\perp^2 \ra}\,\exp\left(-\frac{\bm{p}_{\perp}^2}{\la p_\perp^2 \ra}\right)\;. 
\label{eq:f1D1}
\ea
Note that this is a correct representation of TMDs at tree level; as we mentioned in the Introduction we neglect possible effects coming from resummation of soft gluons.
The collinear distributions $f^{a}_1(x)$ and collinear fragmentation functions $D_{h/a}(z)$ in Eq.~\eqref{eq:f1D1} will follow the usual DGLAP evolution in $Q^2$; we omit the explicit dependence on $Q^2$ in all formulas for simplicity.

Using Eqs.~\eqref{eq:fuu},~\eqref{eq:f1D1}  we obtain
\begin{eqnarray}
F_{UU,T}(x,z,y,P_{hT}) &=& x \sum_{a=q,\bar q} e_a^2 f^a_1(x) D_{h/a}(z) \frac{1}{\pi \la P_{hT}^2 \ra} \exp{\left(-\frac{\bm{P}_{hT}^2}{\la P_{hT}^2 \ra} \right)}\; ,
\label{eq:FUU}
\end{eqnarray}
where
\ba
\la P_{hT}^2 \ra = \la p_{\perp}^2 \ra + z^2 \la k_{\perp}^2 \ra .
\ea
Experimentally one can access $F_{UU,T}(x,z,y,P_{hT})$ by measuring unpolarized multiplicities of hadrons (pions, and kaons) in SIDIS.
Recent analysis of unpolarized multiplicity data of the HERMES Collaboration \cite{Airapetian:2012ki} is presented in Ref.~\cite{Signori:2013mda} and analysis of  data  of the HERMES \cite{Airapetian:2012ki} and COMPASS \cite{Adolph:2013stb} collaborations is presented in Ref.~\cite{Anselmino:2013lza}. 
Note that in principle, the widths of distribution and fragmentation functions $\la k_\perp^2 \ra$ and $\la p_\perp^2 \ra$ can be flavor dependent
and can be functions of $x$ and $z$ correspondingly; however, for the sake of the present analysis such dependencies are not
very important and we will use a more simplified model \cite{Anselmino:2006rv} in which
$\la k_\perp^2 \ra = 0.25$ GeV$^2$ and $\la p_\perp^2 \ra = 0.2$ GeV$^2$. In fact these values were used in extractions ~\cite{Anselmino:2008jk, Anselmino:2007fs, Anselmino:2013vqa} of the Collins fragmentation
functions that we will utilize in this paper.

\subsection{Polarized structure function, $F_{UT}^{\sin(3\phi_h-\phi_S)}$\label{sectionIIB}}
\noindent
The partonic interpretation \cite{Mulders:1995dh,Bacchetta:2006tn,Anselmino:2011ch} of the structure function $F_{UT}^{\sin(3\phi_h -\phi_S)}$ involves the pretzelosity distribution ($h_{1T}^{\perp } $) and the so-called
Collins fragmentation function ($ H_{1}^{\perp }$):
\be
F_{UT}^{ \sin(3\phi_h -\phi_S) }\!
 = {\cal C}\biggl[
   \frac{-2\,\bigl( \hat{\bm{h}}\cdott {\bm{p}_\perp} \bigr)
   \,\bigl( \hat{\bm{h}}\cdott \bm{k}_\perp \bigr)
   -\bm{k}_\perp^2 \bigl( \hat{\bm{h}}\cdott \bm{p}_\perp \bigr) + 4 \bigl( \hat{\bm{h}}\cdott \bm{k}_\perp \bigr)^2 \bigl( \hat{\bm{h}}\cdott\bm{p}_\perp \bigr)
    }{2 M^2 M_h z}
    h_{1T}^{\perp } H_{1}^{\perp }\biggr],
\ee
where $\hat{\bm{h}} \equiv \bm{P}_{h T}/|\bm{P}_{hT}|$.

There exists a positivity bound \cite{Bacchetta:1999kz} for  $h^{\perp a}_{1T}$ 
\be
\frac{\bm{k}_\perp^2}{2 M^2}\left| h^{\perp a}_{1T}(x,k_{\perp}^2) \right|   \le \frac{1}{2}(f_{1}^{a}(x,k_{\perp}^2) - g_{1}^{a}(x,k_{\perp}^2)) \, .
\label{eq:bound}
\ee 
We assume Gaussian form for   $ g_{1}^{q}(x,k_{\perp}^2)$ :
\ba
g_{1}^{a}(x,k_{\perp}^2) = g_{1}^{a}(x)\,\frac{1}{\pi \la k_\perp^2 \ra}\,\exp\left(-\frac{\mathbf{k}_{\perp}^2}{\la k_\perp^2 \ra}\right) \;, 
\label{eq:f1g1}
\ea
where the width $\la k_\perp^2 \ra = 0.25$ (GeV$^2$) is the same as for $f^a_1$. The widths could in principle be different;however, given the precision of the experimental data, such an approximation is a reasonable one. The helicity distributions $g_1(x)$ are taken
from Ref.~\cite{deFlorian:2009vb}, and parton distributions $f_1(x)$ are the GRV98LO PDF set~\cite{Gluck:1998xa}.

We assume the following form of $h_{1T}^{\perp a}$, that preserves  the positivity bound of Eq.~\eqref{eq:bound}:
\ba
h_{1T}^{\perp a}(x,k_{\perp}^2) = \frac{M^2}{M_T^2} e^{-k_\perp^2/M_T^2} h^{\perp a}_{1T}(x) \frac{1}{\pi \la k_\perp^2 \ra}\,\exp\left(-\frac{\bm{k}_{\perp}^2}{\la k_\perp^2 \ra}\right)\;,
\label{eq:h1Tperp}
\ea
where
\begin{align}
h^{\perp a}_{1T}(x) = e  {\cal N}^a(x) (f_{1}^{a}(x) - g_{1}^{a}(x)), \label{eq:hx_par}\\
{\cal N}^a(x) = N^{a} x^{\alpha} (1-x)^{\beta} \frac{(\alpha + \beta)^{\alpha + \beta}}{\alpha^{\alpha} \beta^{\beta}}\, ,\nonumber  
\end{align}
where ${N}^a$, $\alpha$, $\beta$, and $M_T$ will be fitted to data, with $-1\le {N}^{a} \le 1$.

\noindent We use Eq.~(\ref{pdf_moments_our_main}) to calculate the first moment of $ h_{1T}^{\perp a}(x,p_{T}^2)$
of Eq.~(\ref{eq:h1Tperp}) and obtain:
\be
h_{1T}^{\perp (1) a}(x) = \frac{h^{\perp a}_{1T}(x) M_T^2 \la k_\perp^2 \ra}{2 (M_T^2 + \la k_\perp^2 \ra)^2} \, .
\ee
The parametrization of  Collins fragmentation function $H_{1}^{\perp q}$ is taken from Refs.~\cite{Anselmino:2008jk, Anselmino:2007fs, Anselmino:2013vqa}:
 \ba
 \label{tr-funct} 
H_{1 h/a}^{\perp}(z, p_\perp) = \frac{z M_h}{2 p_\perp} \Delta^N \! D_{h/a^\uparrow}(z,p_\perp) &=&  \frac{z M_h}{M_C} e^{-p_{\perp}^2/M_C^2} H_{1 h/a}^{\perp}(z)\frac{1}{\pi \la p_\perp^2 \ra}\,\exp\left(-\frac{\bm{p}_{\perp}^2}{\la p_\perp^2 \ra}\right)\;,
\label{coll-funct}
 \ea
 with
\begin{align}
H_{1 h/a}^{\perp}(z) = \sqrt{2 e} {\cal N}^{\C}_a(z) D_{h/a}(z), \nonumber \\
{\cal N}^{\C}_a(z)= N^{\C}_a \, z^{\gamma} (1-z)^{\delta} \,
\frac{(\gamma + \delta)^{(\gamma +\delta)}}
{\gamma^{\gamma} \delta^{\delta}},
\label{eq:coll_par}
 \end{align}
where  $-1 \le N^{\C}_a \le 1$ and $\la p_\perp^2 \ra = 0.2$ (GeV$^2$). The  fragmentation functions (FF)  $D_{h/a}(z)$ are from the 
DSS LO fragmentation function set~\cite{deFlorian:2007aj}. Notice that with these choices  
  the Collins fragmentation function automatically obeys its
proper positivity bound \cite{Bacchetta:1999kz}. In the fits we use the parameters of Collins FF obtained in Ref.~\cite{Anselmino:2013vqa}. Note that as in Ref.~\cite{Anselmino:2013vqa} we use two 
Collins fragmentation functions, {\it favored} and {\it unfavored} ones  (see Ref.~\cite{Anselmino:2013vqa} for details on implementation), and corresponding parameters ${N}^{\C}_a$ are then  ${N}^{\C}_{fav}$ and ${N}^{\C}_{unfav}$.

According to Eq.\eqref{half_moments_our_main} we obtain the following expression for the half moment of 
the Collins fragmentation function: 
\ba
H_{1 h/a}^{\perp (1/2)}(z) = \frac{H_{1 h/a}^{\perp}(z) M_C^2}{4}\sqrt{\frac{\pi \la p_\perp^2 \ra}{(M_C^2+\la p_\perp^2 \ra)^3}}\; .
\ea 
We also define the following variables:
\begin{align}
\la k_{\perp}^2 \ra_T  = \frac{\la k_\perp^2 \ra M_T^2}{\la k_\perp^2 \ra + M_T^2} ,\;
\la p_{\perp}^2 \ra_C = \frac{\la p_\perp^2 \ra M_C^2 }{\la p_\perp^2 \ra + M_C^2} ,  \nonumber \\
\la P_{hT}^2 \ra_{CT} = \la p_{\perp}^2 \ra_C + z^2 \la k_{\perp}^2 \ra_T  \, .
\end{align}
The polarized structure function  $F_{UT}^{ \sin(3\phi_h -\phi_S)}$ can be readily computed and reads   (see also Ref.~\cite{Anselmino:2011ch})
\ba
F_{UT}^{\sin(3\phi_h-\phi_S)}(x,z,y,P_{hT}) &=& \frac{x z^2  P_{hT}^{3}}{2} \sum_{a=q,\bar q} e_a^2 
h_{1T}^{\perp (1) a}(x) H_{1 h/a}^{\perp (1/2)}(z)
\frac{{\cal C}}{ \pi \la P_{hT}^2 \ra_{CT}^4} e^{-P_{hT}^2/\la P_{hT}^2 \ra_{CT}} \; ,
\label{numerator_new}
\ea
where
\ba
{\cal C} = 8 \la k_{\perp}^2 \ra_T\sqrt{\frac{\la p_{\perp}^2 \ra_C}{\pi}}\, .
\ea
One can see from Eq.~\eqref{oam} that under these assumptions on relation of pretzelosity to OAM we obtain
\ba
F_{UT}^{\sin(3\phi_h-\phi_S)} \propto \sum_{q} e_q^2 {\cal L}_{z}^{q}(x) \; .
\ea
Note that the model relation of Eq.\eqref{oam} is found to be valid only for quarks and not for antiquarks, in this study we will neglect potential contributions from anti-quarks.

The experimental data are presented as sets of projections on $x$, $z$, and ${P}_{hT}$. In fact the three data sets  are a projection of the same data set and not independent; thus, in principle we
should not include all sets in the fit. However provided that
the projections are at different average values of $x$, $z$, and ${P}_{hT}$ we do gain sensitivity to distribution and fragmentation functions
if we include simultaneously all three data sets. In the following we will assume them to be independent and include them into our $\chi^2$ analysis. However, it would be clearly beneficial for the phenomenological analysis if experimental data were presented in a simultaneous 4-D $x$, $z$, $y$, ${P}_{hT}$ binning.
For  the asymmetry as a function of $x,z$ we are using our result of Eq.~\eqref{eq:final}, in particular, the value of the experimental point's  $\langle {P}_{hT} \rangle$. 

We also include a simplified scale dependence in the asymmetry  by using $Q^2$ in the corresponding collinear distribution.
The collinear quantities $h_{1T}^{\perp (1) a}(x)$ and $H_{1 h/a}^{\perp (1/2)}(z)$ in general will be related to the so-called twist-3 matrix elements related to multiparton correlations. Such matrix elements have a nontrivial QCD evolution (for example, Ref.~\cite{Kang:2008wv}). The complete solutions of evolution equations are currently unknown and we substitute them by DGLAP evolution of corresponding collinear distributions in the parametrizations of Eqs.~(\ref{eq:hx_par}),(\ref{eq:coll_par}).
 
For completeness we also give  results for ${P}_{hT}$  integrated asymmetry
\begin{align}
 A_{UT}^{\sin(3\phi_h-\phi_S)}(x,z,y) =  \frac{  \frac{\alpha^2}{s x^2 y^2} 2 \left(1 - y\right)\, 
\int d P_{hT} P_{hT} d \phi_h \; F_{UT}^{\sin(3\phi_h-\phi_S)}}{\frac{\alpha^2}{s x^2 y^2} \left(1+(1-y)^2\right) \int d P_{hT} P_{hT} d \phi_h \; F_{UU ,T}}\;.
\label{eq:final1}
\end{align}
Using Eqs.~\eqref{eq:FUU},\eqref{numerator_new} we obtain
\begin{eqnarray}
\int d P_{hT} P_{hT} d \phi_h \; F_{UU,T}(x,z,y,P_{hT}) &=&  x\,  \sum_{a=q,\bar q} e_a^2 f_1^a(x) D_{h/a}(z)    \; , \nonumber \\
 \int d P_{hT} P_{hT} d \phi_h \; F_{UT}^{\sin(3\phi_h-\phi_S)} (x,z,y,P_{hT}) &=&   \frac{x z^2 }{2} \sum_{a=q,\bar q} e_a^2 
h_{1T}^{\perp (1) a}(x) H_{1 h/a}^{\perp (1/2)}(z)
\frac{3 \, {\cal C} \sqrt{\pi}}{ 4\, \la P_{hT}^2 \ra_{CT}^{3/2}}   \; .
\label{eq:FUTintegrated}
\end{eqnarray}
${P}_{hT}$  integrated asymmetry was used for comparison with experimental results in Refs.~\cite{She:2009jq,Avakian:2008dz}. We checked explicitly that results 
 of fitting with average values of  $\langle {P}_{hT} \rangle$ and ${P}_{hT}$  integrated ones are consistent with each other. 
 
 In the following we will use values of $\langle x \rangle$, $\langle y \rangle$, $\langle z \rangle$, $\langle {P}_{hT} \rangle$ in each experimental point  to estimate the asymmetry using Eq.~\eqref{numerator_new}.

\section{Null signal hypothesis\label{sectionIII}}
\noindent
Before proceeding to the phenomenology of pretzelosity, let us try to understand if the experimental data are compatible with a
null hypothesis. We calculate the probability that $h_{1T}^{\perp a}(x, p_T^2) \equiv 0$ or $F_{UT}^{\sin(3\phi_h-\phi_S)}\equiv 0$.

We calculate thus the value of
\ba
\chi^2_0 = \sum_{n=1}^{N_{data}} \left(\frac{F_{UT}^{\sin(3\phi_h-\phi_S)}}{\Delta F_{UT}^{\sin(3\phi_h-\phi_S)}}\right)^2\; ,
\ea
where the experimental error is $\Delta F = \sqrt{{\Delta F}_{sys}^2 + {\Delta F}_{stat}^2}$.

Results are presented in Table.~\ref{tableI}. We see that the total value of $\chi^2 =  163.48$ for $N_{data} = 175$.

\begin{table}[h]
\begin{tabular}{l l l l l c c}
Experiment & Hadron & Target &  Dependence & \# ndata & $\chi^2$ & $\chi^2/{ndata}$ \\
\hline
COMPASS \cite{Parsamyan:2007ju} &$h^+$ &LiD& $x$  & 9 &   2.12   & 0.23 \\
COMPASS \cite{Parsamyan:2007ju} &$h^-$ &LiD&  $x$  & 9 &   5.66  & 0.62 \\  
COMPASS \cite{Parsamyan:2007ju} &$h^+$ &LiD&  $z$  & 8 &  15.45   & 1.93 \\ 
COMPASS \cite{Parsamyan:2007ju} &$h^-$ &LiD&  $z$  & 8 &   3.64  & 0.45 \\ 
COMPASS \cite{Parsamyan:2007ju} &$h^+$ &LiD&  $P_{h T}$  & 9 &  10.05   & 1.11  \\ 
COMPASS \cite{Parsamyan:2007ju} &$h^-$ &LiD&  $P_{h T}$  & 9 &  10.46   & 1.16  \\ 
\hline
COMPASS \cite{Parsamyan:2013fia} &$h^+$   &NH$_3$&$x$  & 9 &   11.28   & 1.25 \\
COMPASS \cite{Parsamyan:2013fia} &$h^-$   &NH$_3$&$x$  & 9 &  4.30   & 0.48 \\  
COMPASS \cite{Parsamyan:2013fia} &$h^+$  &NH$_3$&$z$  & 8 &   13.76  & 1.72 \\ 
COMPASS \cite{Parsamyan:2013fia} &$h^-$   &NH$_3$&$z$  & 8 &  1.69   & 0.21 \\ 
COMPASS \cite{Parsamyan:2013fia} &$h^+$  &NH$_3$&$P_{h T}$  & 9 &    11.12 & 1.24  \\ 
COMPASS \cite{Parsamyan:2013fia} &$h^-$   &NH$_3$&$P_{h T}$  & 9 &    8.07  & 0.90  \\ 
\hline
HERMES \cite{Diefenthaler:2010zz,Schnell:2010zza,Pappalardo:2010zz} &$\pi^0$ &H&  $x$  & 7 &  12.29  & 1.76 \\ 
HERMES \cite{Diefenthaler:2010zz,Schnell:2010zza,Pappalardo:2010zz} &$\pi^+$ &H&  $x$  & 7 &   2.99   & 0.43 \\
HERMES \cite{Diefenthaler:2010zz,Schnell:2010zza,Pappalardo:2010zz} &$\pi^-$ &H&  $x$  & 7 &  10.12   & 1.45 \\ 
HERMES \cite{Diefenthaler:2010zz,Schnell:2010zza,Pappalardo:2010zz} &$\pi^0$ &H&  $z$  & 7 &    2.24 & 0.32 \\ 
HERMES \cite{Diefenthaler:2010zz,Schnell:2010zza,Pappalardo:2010zz} &$\pi^+$ &H&  $z$  & 7 &   5.14  & 0.73 \\ 
HERMES \cite{Diefenthaler:2010zz,Schnell:2010zza,Pappalardo:2010zz} &$\pi^-$ &H&  $z$  & 7 &  3.68   & 0.52 \\
HERMES \cite{Diefenthaler:2010zz,Schnell:2010zza,Pappalardo:2010zz} &$\pi^0$ &H&  $P_{h T}$  & 7 &    5.74 & 0.82 \\ 
HERMES \cite{Diefenthaler:2010zz,Schnell:2010zza,Pappalardo:2010zz} &$\pi^+$ &H&  $P_{h T}$  & 7 &    4.92 & 0.70  \\ 
HERMES \cite{Diefenthaler:2010zz,Schnell:2010zza,Pappalardo:2010zz} &$\pi^-$ &H&  $P_{h T}$  & 7 &    12.89  & 1.84  \\ 
  \hline
JLAB \cite{Zhang:2013dow} &$\pi^+$ &$^3$He&  $x$  & 4 &   4.35   & 1.19 \\
JLAB \cite{Zhang:2013dow} &$\pi^-$ &$^3$He&  $x$  & 4 &  1.52   & 0.41 \\ 
\hline
 & & & & 175 & 163.48 & 0.93 \\
\end{tabular}
\caption{Results of the analysis of the null hypothesis.}
\label{tableI}
\end{table}
The goodness of this fit for a given $\chi^2$ is normally calculated as
$P(\chi^2,n_{d.o.f.})$, the integral of the probability distribution in
$\chi^2$ for $n_{d.o.f.}$ degrees of freedom, integrated from the observed minimum $\chi^2_0$
to infinity:
\ba
P(\chi^2_0,n_{d.o.f.}) = 1 - \int_{0}^{\chi^2_0} d \chi^2 \frac{1}{2 \Gamma(n_{d.o.f.}/2)}\left( \frac{\chi^2}{2} \right)^{n_{d.o.f.}/2 -1} \exp \left[-{\frac{\chi^2}{2}}\right]
\ea

We obtain $P(163.48,175) = 72\%$; i.e., there is a good chance that the quark-charge weighted sums over pretzelosity are zero or, in particular, all $h_{1T}^{\perp a}(x, p_T^2) = 0$.
However we note that $F_{UT}^{\sin(3\phi_h-\phi_S)} \propto z^2  P_{h\perp}^{3}$ and thus asymmetry is suppressed by an additional factor
of $z  P_{h\perp}^{2}$ with respect to Collins asymmetries, which were experimentally  found to be nonzero. The latter
do not usually exceed 10\%. Assuming that $\langle z \rangle \sim 0.5$, $\langle P_{h\perp} \rangle \sim 0.5$ GeV, we conclude that even if
$h_{1T}^{\perp a}(x, p_T^2)$ is of the same magnitude as the transversity distribution (that couples to Collins FF and generates Collins SSA), one would expect $F_{UT}^{\sin(3\phi_h-\phi_S)}\sim 1.5$\% at most. In fact the maximum asymmetry due to pretzelosity was estimated to be 
of the order of $\sim 5$\% in Ref.~\cite{Avakian:2008dz}. One can see that preliminary HERMES and COMPASS data are indeed in the range $|A_{UT}^{\sin(3\phi_h-\phi_S)}|\lesssim 2$\%; thus, we will attempt fitting the data. We emphasize that future JLab 12 \cite{Dudek:2012vr} data are going to be of extreme importance for exploring pretzelosity TMD in the valence quark region.

\section{Phenomenology\label{sectionIV}}
\noindent
In our analysis we are going to fit the unknown parameters for pretzelosity distributions.
The precision of the experimental data is quite low, thus we are going to set $\alpha$, $\beta$, and $M_T$ to be flavor independent.
We saw in the previous section that the data are compatible with zero pretzelosity and experimental errors are big. Therefore one is not able at present to determine all parameters and we will fix $\beta = 2$ as pretzelosity is expected \cite{Avakian:2007xa,Brodsky:2006hj,Burkardt:2007rv} to be suppressed  by $(1-x)^2$ with respect to an unpolarized distribution. We also assume $\alpha_u = \alpha_d \equiv \alpha$.
Thus we are going to fit four parameters: $N_u, N_d, \alpha$, and $M_T^2$.

We will fit $h^+$ and $h^-$ data and on effective deuteron (LiD) \cite{Parsamyan:2007ju} and proton (NH$_3$) \cite{Parsamyan:2013fia}  targets  from the COMPASS Collaboration,  $\pi^0$, $\pi^+$, and $\pi^-$ data on proton (H) target  from preliminary HERMES \cite{Diefenthaler:2010zz,Schnell:2010zza,Pappalardo:2010zz},   and JLab 6 data \cite{Zhang:2013dow} on an effective neutron ($^3$He) target.

Note that the COMPASS data  \cite{Parsamyan:2007ju,Parsamyan:2013fia} are presented in the following way:
\begin{align}
{A_{UT}^{\sin(3\phi_h-\phi_S)}}_{COMPASS} = \frac{A_{UT}^{\sin(3\phi_h-\phi_S)}}{D_{NN}  (\la y \ra)}  \; ,
\end{align}
where 
\begin{align}
D_{NN} (\la y \ra) =\frac{2(1-\la y \ra)}{1+(1-\la y \ra)^2}
\end{align}
and $A_{UT}^{\sin(3\phi_h-\phi_S)}$ is from Eq.\eqref{eq:final}. In our fitting procedure we take $D_{NN}(\la y\ra)$ into account and use experimental values of 
$\la y\ra$ for each bin. The value of $Q^2$ is always set by the experiment and varies from bin to bin.

  Parameters of the Collins fragmentation function are taken from Ref.~\cite{Anselmino:2013vqa} and presented in Table~\ref{collins}.

\begin{table}[h]
\begin{tabular}{l l}
 ${N}^{\C}_{fav} =$ $0.49^{+0.2}_{-0.18} $ & ${N}^{\C}_{unfav} = $ $-1^{+0.38}_{-0}$\\
 $\gamma = $ $1.06^{+0.45}_{-0.32}$ & $\delta = $ $0.07^{+0.42}_{-0.07}$\\
 $M_C^2 = $ $1.50^{+2.00}_{-1.12}$  (GeV$^2$)    \\
%
\end{tabular}
\caption{Parameters of Collins FF (Ref.~\cite{Anselmino:2013vqa})}
\label{collins}
\end{table}

The resulting parameters after the fit are presented in Table~\ref{fitparI} and partial values of $\chi^2$ are presented in Table.~\ref{tableII}. 
One can easily see that the modern experimental data do not allow for a precise extraction of pretzelosity as the errors reported in
Table.~\ref{fitparI} are quite big. However,
one notes that positive values for $N_u$ and negative for $N_d$ are preferred by the data. 

\begin{figure*}[htb]
  \begin{tabular}{c@{\hspace*{2cm}}c}
     \includegraphics[width=8cm]{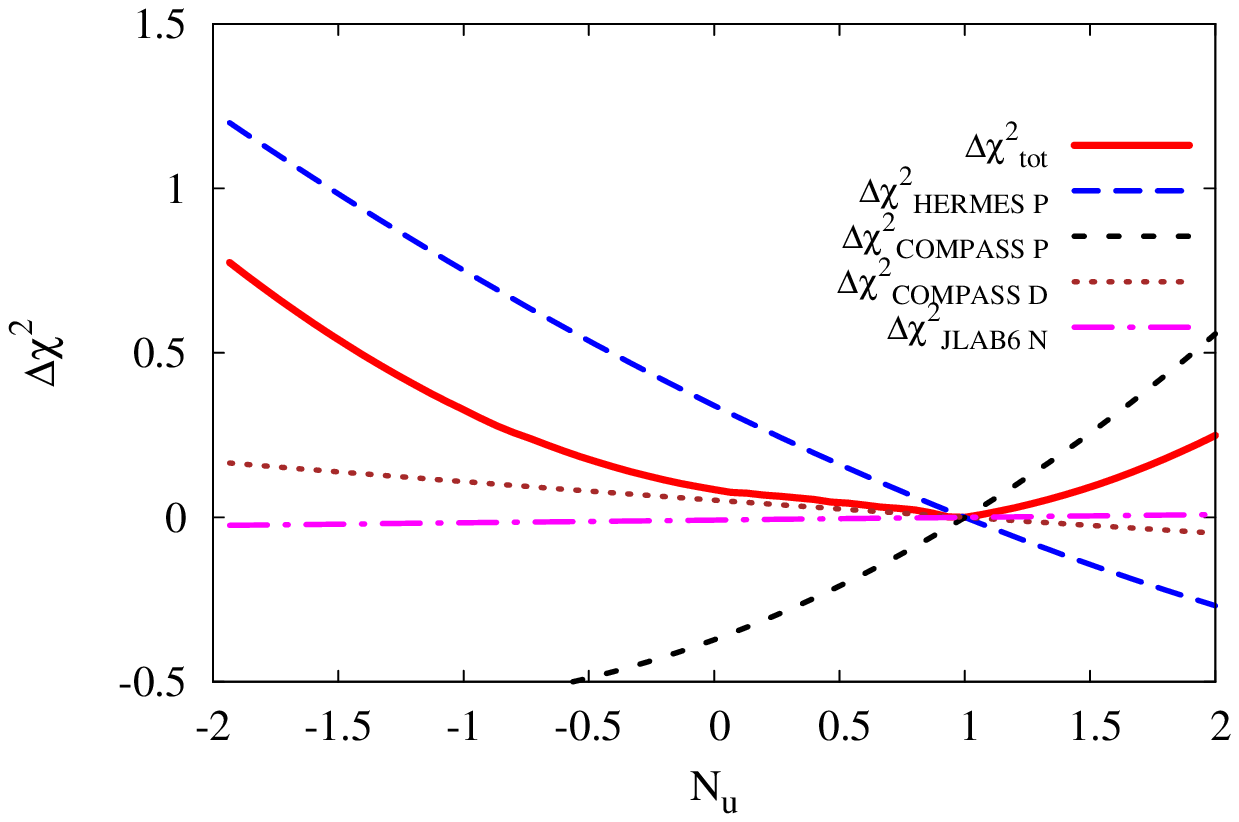}
    &
    \includegraphics[width=8cm]{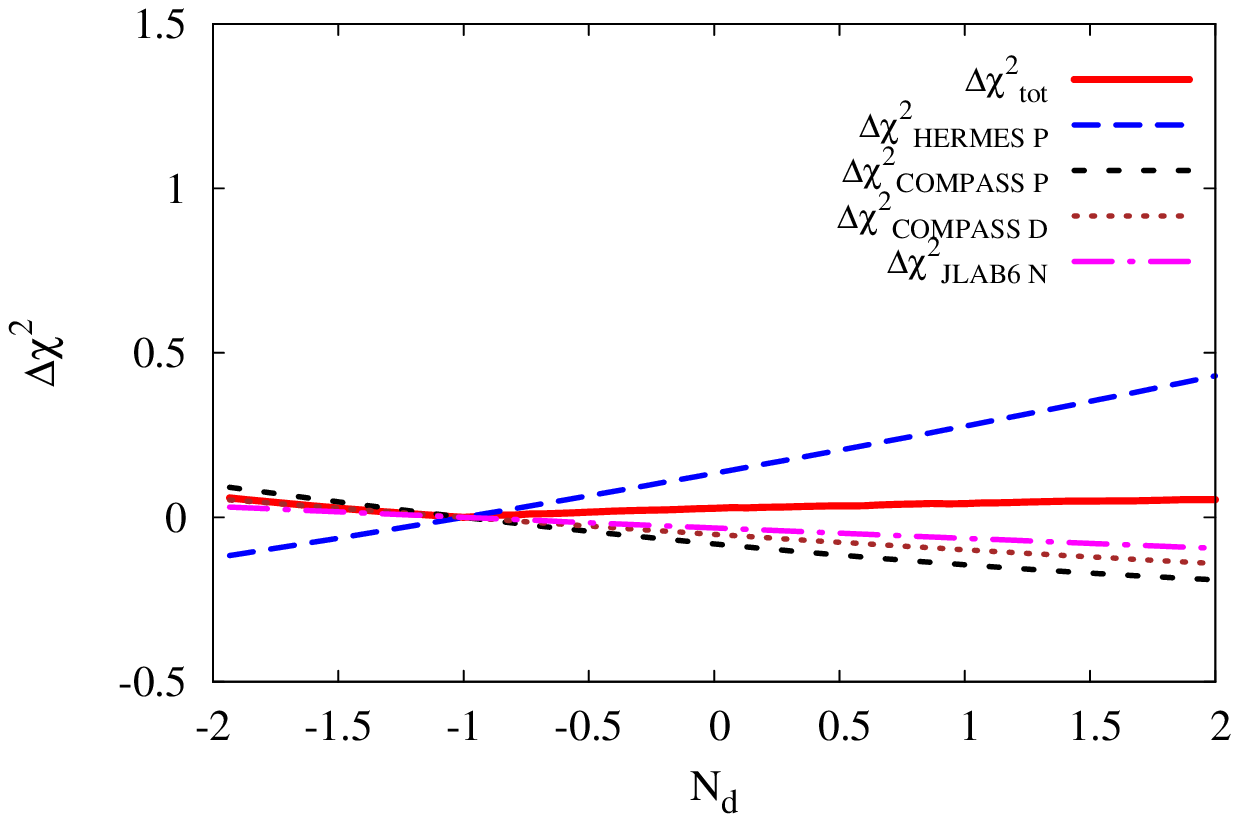}
  \\
  (a) & (b)
  \end{tabular}     
\caption{$\Delta \chi^2$ as a function of $N_{u}$ (a) and $N_{d}$ (b)  for total $\chi^2$ and preliminary COMPASS \cite{Parsamyan:2007ju},  \cite{Parsamyan:2013fia},    HERMES \cite{Diefenthaler:2010zz,Schnell:2010zza,Pappalardo:2010zz},  and JLab  \cite{Zhang:2013dow}  separately.}
\label{fig:chi2}
\end{figure*}

In order to check which values of parameters are preferred by individual data sets we vary $N_u \in [-2,2]$ and $N_d \in [-2,2]$ and fix all other parameters to the best fit values.
We calculate the total $\chi^2$ and partial values of $\chi^2_{\rm COMPASS\; D}$, $\chi^2_{\rm COMPASS \; P}$,    $\chi^2_{\rm HERMES\; P}$, $\chi^2_{\rm JLAB\; N}$ coming from data sets COMPASS  \cite{Parsamyan:2007ju,Parsamyan:2013fia}, \cite{Parsamyan:2013fia},   HERMES \cite{Diefenthaler:2010zz,Schnell:2010zza,Pappalardo:2010zz},  and JLab \cite{Zhang:2013dow} . We then plot
$\Delta \chi^2 \equiv \chi^2 - \chi^2_{min}$   as a function of $N_u$ in Fig.~\ref{fig:chi2} (a) and as 
a function of $N_d$ in Fig.~\ref{fig:chi2} (b). Here $\chi^2_{min}$ corresponds to the best fit. The point where all curves intersect corresponds to the best fit value for $N_{u,d}$.

One can see from Fig.~\ref{fig:chi2} that preliminary HERMES data prefer positive values for $N_u$ and negative values for $N_d$ and this tendency is the most prominent. COMPASS data, however, prefer negative values for $N_u$ and positive values for $N_d$. The fit of all data sets in turn follows preference to  positive values for $N_u$ and negative values for $N_d$. The major part of the data comes from the proton target; thus as expected, we have a better determination of $N_u$ due to up-quark dominance and $\Delta \chi^2$ in fact is the biggest in this case (Fig.~\ref{fig:chi2} (a)). We also expect 
that parameter $N_d$ will be determined with bigger uncertainty (Fig.~\ref{fig:chi2} (b)). One can also see from Fig.~\ref{fig:chi2} that we cannot establish that pretzelosity does not violate positivity bounds; in fact, values beyond region $[-1,1]$ are also possible. We performed a study of possible positivity bound Eq.~\eqref{eq:bound} violation by pretzelosity and found no evidence 
of such a violation in existing experimental data, in other words the fit does not yield values of $N_u$, $N_d$ violating the positivity bound if these parameters are
allowed to vary in a bigger region.

The errors of extraction are estimated using the Monte Carlo method from Ref.~\cite{Anselmino:2008sga}.  We generate 500 sets of parameters ${\bf a}_j = \{\alpha_j,N_{u\, j},N_{d\, j},M_{T\, j}^{2}\}$ that satisfy
\be
\chi^2({\bf a}_j) \le \chi^2_{min} + \Delta \chi^2\, ,
\label{condition}
\ee
where $\Delta \chi^2 = 9.72$ that corresponds to  $P=95.45\%$ of coverage probability for four parameters. Those parameter sets are then used to estimate the errors.

\begin{table}[htb]
\begin{tabular}{l c l l c l}
\hline
$\alpha$ &=& $2.5\pm1.5$ & $\beta$ &=& $2$ fixed \\
 $N_{u}$ &=& $1 \pm 1.4$ & $N_{d}$ &=& $-1 \pm 1.3$\\
 $M_T^2$ &=& $0.18 \pm  0.7$&\multicolumn{3}{l}{(GeV$^2$)}\\ 
\hline
\multicolumn{3}{l}{$\chi_{min}^2 =  163.33$} & \multicolumn{3}{l}{$\chi^2_{min}/{n.d.o.f}=0.95$}  \\
\end{tabular}
\caption{Fitted parameters of the pretzelosity quark distributions.}
\label{fitparI}
\end{table}

\begin{table}[htb]
\begin{tabular}{l l l c c c c}
Experiment & Hadron & Target &  Dependence & \# ndata & $\chi^2$ &
$\chi^2/{ndata}$ \\
\hline
COMPASS \cite{Parsamyan:2007ju} &$h^+$ &LiD& $x$  & 9 &   2.11   & 0.23 \\
COMPASS \cite{Parsamyan:2007ju} &$h^-$ &LiD&  $x$  & 9 &   5.68   & 0.63 \\
COMPASS \cite{Parsamyan:2007ju} &$h^+$ &LiD&  $z$  & 8 &  15.45   & 1.93 \\
COMPASS \cite{Parsamyan:2007ju} &$h^-$ &LiD&  $z$  & 8 &   3.63  & 0.45 \\
COMPASS \cite{Parsamyan:2007ju} &$h^+$ &LiD&  $P_{h T}$ & 9 &  10.05   &
1.12  \\
COMPASS \cite{Parsamyan:2007ju} &$h^-$ &LiD&  $P_{h T}$ & 9 &  10.46   &
1.16  \\
\hline
COMPASS \cite{Parsamyan:2013fia} &$h^+$   &NH$_3$&$x$ & 9 &   11.22   &
1.25 \\
COMPASS \cite{Parsamyan:2013fia} &$h^-$   &NH$_3$&$x$ & 9 &  4.21   &
0.47 \\
COMPASS \cite{Parsamyan:2013fia} &$h^+$  &NH$_3$&$z$ & 8 &   13.92  &
1.74 \\
COMPASS \cite{Parsamyan:2013fia} &$h^-$   &NH$_3$&$z$ & 8 &  1.67  &
0.20 \\
COMPASS \cite{Parsamyan:2013fia} &$h^+$  &NH$_3$&$P_{h T}$  & 9 &
11.23 & 1.25  \\
COMPASS \cite{Parsamyan:2013fia} &$h^-$   &NH$_3$&$P_{h T}$  & 9 &
8.04  & 0.89  \\
\hline
HERMES \cite{Diefenthaler:2010zz,Schnell:2010zza,Pappalardo:2010zz} &$\pi^0$ &H&  $x$  & 7 & 12.27  & 1.75 \\
HERMES \cite{Diefenthaler:2010zz,Schnell:2010zza,Pappalardo:2010zz} &$\pi^+$ &H&  $x$  & 7 & 3.05   & 0.44 \\
HERMES \cite{Diefenthaler:2010zz,Schnell:2010zza,Pappalardo:2010zz} &$\pi^-$ &H&  $x$  & 7 & 10.06   & 1.44 \\
HERMES \cite{Diefenthaler:2010zz,Schnell:2010zza,Pappalardo:2010zz} &$\pi^0$ &H&  $z$  & 7 & 2.23 & 0.32 \\
HERMES \cite{Diefenthaler:2010zz,Schnell:2010zza,Pappalardo:2010zz} &$\pi^+$ &H&  $z$  & 7 & 5.08  & 0.73 \\
HERMES \cite{Diefenthaler:2010zz,Schnell:2010zza,Pappalardo:2010zz} &$\pi^-$ &H&  $z$  & 7 & 3.47   & 0.50 \\
HERMES \cite{Diefenthaler:2010zz,Schnell:2010zza,Pappalardo:2010zz} &$\pi^0$ &H&  $P_{h T}$  & 7 &    5.74 & 0.82 \\
HERMES \cite{Diefenthaler:2010zz,Schnell:2010zza,Pappalardo:2010zz} &$\pi^+$ &H&  $P_{h T}$  & 7 &    4.84 & 0.69  \\
HERMES \cite{Diefenthaler:2010zz,Schnell:2010zza,Pappalardo:2010zz} &$\pi^-$ &H&  $P_{h T}$  & 7 &    12.93 & 1.85  \\
  \hline
JLAB \cite{Zhang:2013dow} &$\pi^+$ &$^3$He&  $x$  & 4 &   4.35   & 1.09 \\
JLAB \cite{Zhang:2013dow} &$\pi^-$ &$^3$He&  $x$  & 4 &  1.56   & 0.39 \\
\hline
  & & & & 175 & 163.33 & 0.93 \\
\end{tabular}
\caption{Partial $\chi^2$ values of the best fit.}
\label{tableII}
\end{table}

\begin{figure*}[h]
  \begin{tabular}{c@{\hspace*{-1cm}}c}
    \includegraphics[width=6cm,angle=-90]{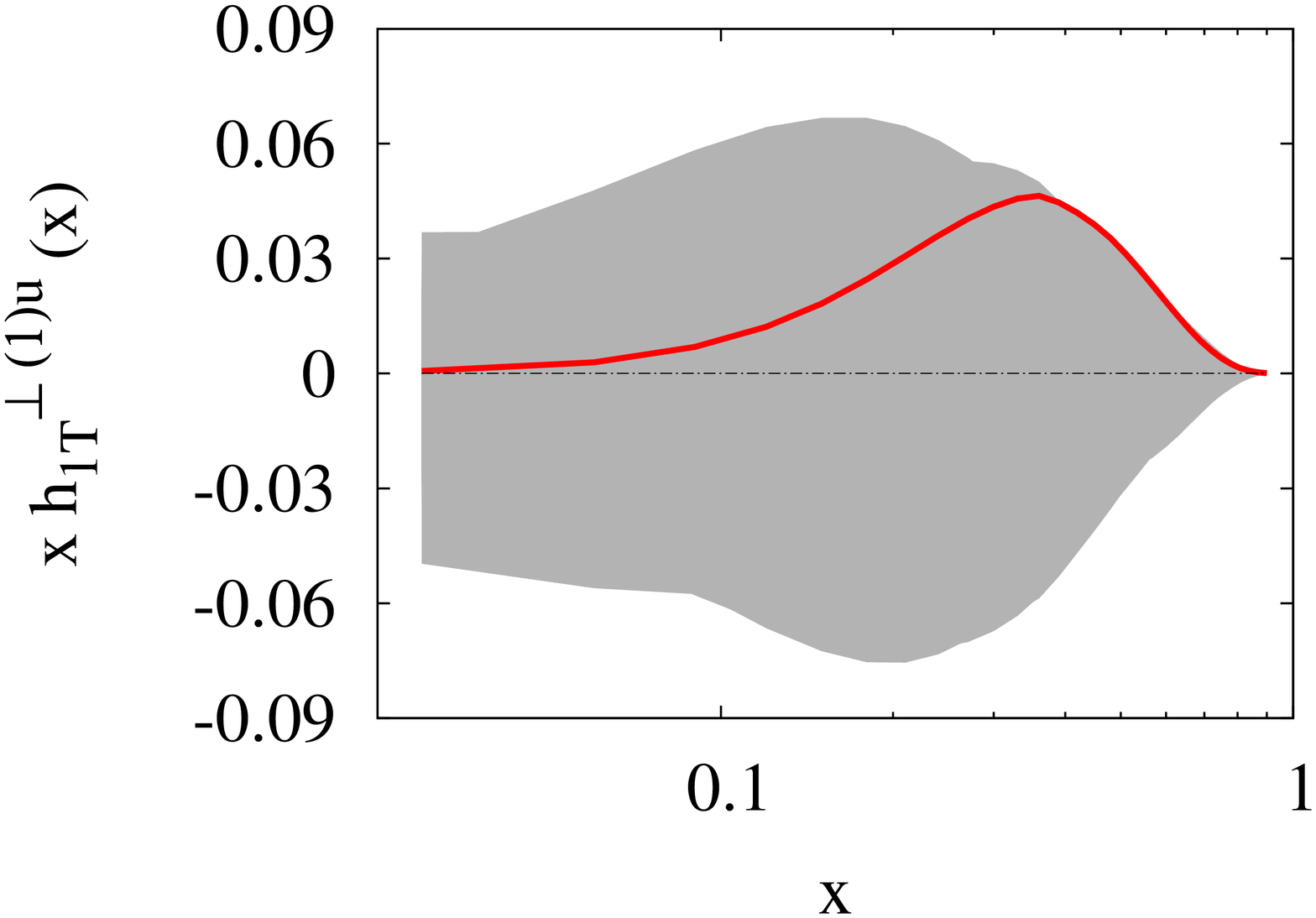}
    &
    \hskip 1cm \includegraphics[width=6cm,angle=-90]{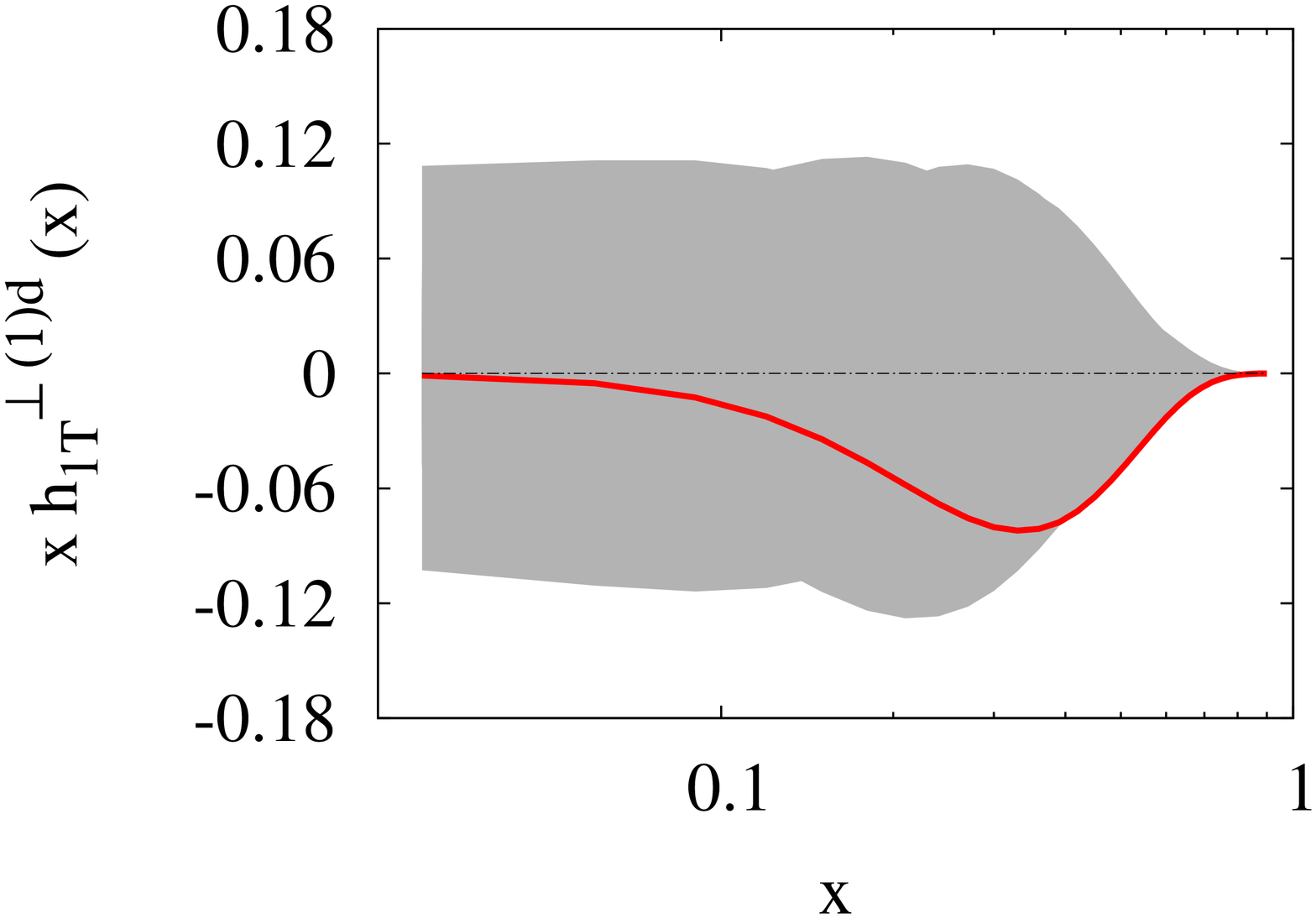}
  \\
  (a) & (b)
  \end{tabular}     
\caption{First moment of the pretzelosity distribution for up (a) and down (b) quarks at $Q^2=2.4$ GeV$^2$. The solid line corresponds to
the best fit and the shadowed region corresponds to the error corridor explained in the text.}
\label{fig:pretzelosity}
\end{figure*}

Resulting pretzelosity is presented in Figure~\ref{fig:pretzelosity}. One can see that resulting pretzelosity has a very large error corridor and diminishes at small $x$. 
Future Jefferson Lab 12 GeV data is going to be crucial for the progress of phenomenology of the pretzelosity distribution as JLab 12 \cite{Dudek:2012vr} data will
explore the  high-$x$ region. Fig.~\ref{fig:pretzelosity} also demonstrates that the best fit indicates positive pretzelosity for up quark and negative pretzelosity for down quark.

\begin{figure*}[h]
\includegraphics[width=12cm]{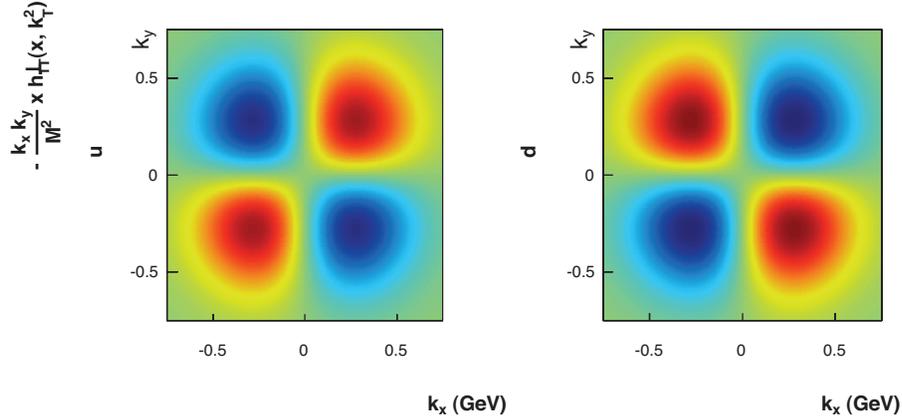}    
\caption{Tomographic slice of the pretzelosity distribution Eq.~\eqref{eq:phi1} at $x = 0.1$ for up and down quarks. Red colors mean positive sign function while blue colors
mean negative sign function.}
\label{fig:pretzelosity3d}
\end{figure*}

We also plot in Fig.~\ref{fig:pretzelosity3d} the quadrupole modulation that corresponds to the pretzelosity distribution with particular choices of $\alpha = 1, \rho = 2$ from Eq.~\eqref{eq:phi}
\begin{align}
   -\frac{1}{2}\frac{{k}_\perp^{1}{k}_\perp^{2}}{M^2} 
  x h_{1T}^{\perp}
(x,{k}_{\perp}^{2})
\label{eq:phi1}
\end{align}
 
One can see from Fig.~\ref{fig:pretzelosity3d} that indeed the quadrupole deformation of distribution is clearly present due to pretzelosity.

\begin{figure*}[htb]
\centering
\includegraphics[width=6cm,angle=-90]{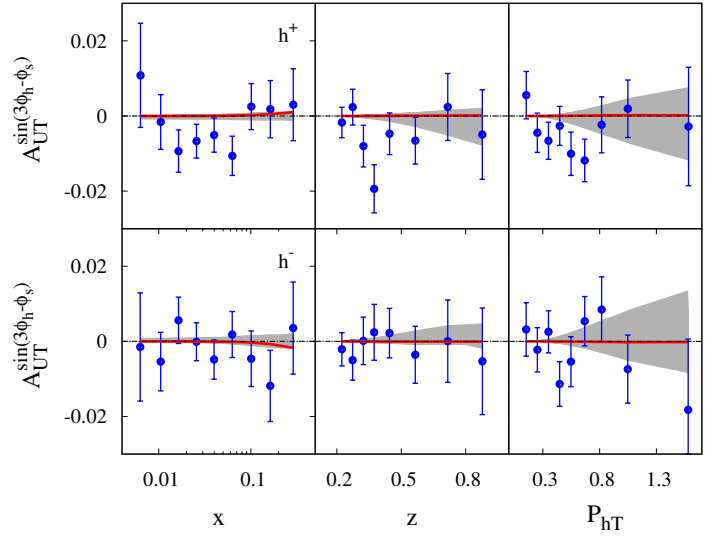}   
\vskip 1.cm
\caption{$A_{UT}^{\sin(3\phi_h-\phi_S)}$ measured by the COMPASS Collaboration on a proton (NH$_3$) target \cite{Parsamyan:2013fia} as a function of $x,z,P_{hT}$ for $h^+$ and $h^-$. The solid line corresponds to
the best fit and the shadowed region corresponds to the error corridor.}
\label{fig:autCOMPASSP}
\end{figure*}

\begin{figure*}[htb]
\centering
\includegraphics[width=6cm,angle=-90]{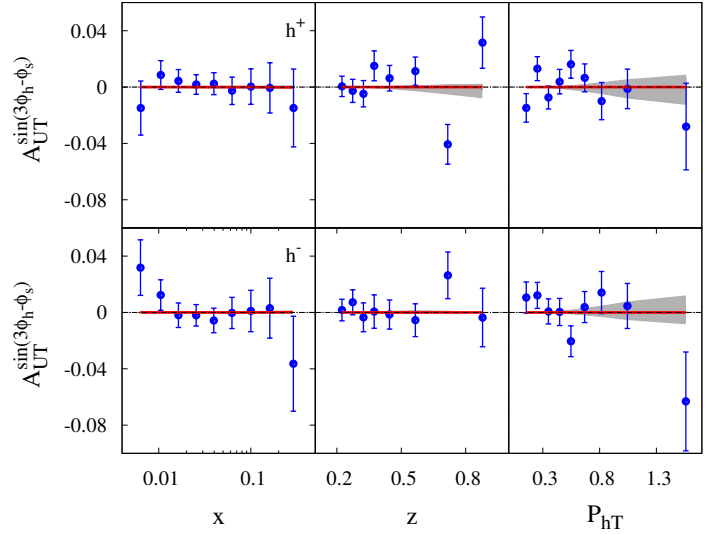}  
\vskip 1.cm
\caption{$A_{UT}^{\sin(3\phi_h-\phi_S)}$  measured by the COMPASS Collaboration on a deuteron (LiD) target \cite{Parsamyan:2007ju} as a function of $x,z,P_{hT}$ for $h^+$  and $h^-$.  The  solid line corresponds to
the best fit and the shadowed region corresponds to the error corridor.}
\label{fig:autCOMPASSD}
\end{figure*}

Results of the description  of  COMPASS \cite{Parsamyan:2013fia,Parsamyan:2007ju}  data on $h^\pm$ production are presented in Fig.~\ref{fig:autCOMPASSP} for a proton (NH$_3$) target and
in Fig.~\ref{fig:autCOMPASSD} for a deuteron (LiD) target. One can see that the expected asymmetry is very small especially for $z$ and $P_{hT}$ dependence, the reason is that
COMPASS $\la x \ra \simeq 0.03$ is quite small and pretzelosity quickly diminishes at small $x$. However, the error corridor is quite large. In addition, cancellation of $u$ and $d$ pretzelosities makes asymmetries on deuteron target vanishing, see Fig.~\ref{fig:autCOMPASSD}. Indeed for $h^+$ production on a deuteron target $F_{UT}^{\sin(3\phi_h-\phi_S)} \propto
 4 (h_{1T}^{\perp u} +  h_{1T}^{\perp d}) H_{1}^{\perp fav} + (h_{1T}^{\perp u} +  h_{1T}^{\perp d}) H_{1}^{\perp unfav} \sim 0$ because our result indicates that $h_{1T}^{\perp u} + h_{1T}^{\perp d}\sim 0$. Overall smallness of asymmetry on the proton target in Fig.~\ref{fig:autCOMPASSP} is due to the suppression factor $z^2 P_{h T}^3$. Our result also indicates that pretzelosity diminishes as $x$ becomes smaller; thus, we have almost vanishing results for small values of $x$. We cannot of course exclude possible contribution from sea quarks or bigger values of pretzelosity in the small-$x$ region. Note that our results are scaled 
by $D_{NN}$ in order to be compared to the COMPASS data . \\

\begin{figure*}[htb]
\includegraphics[width=6cm,angle=-90]{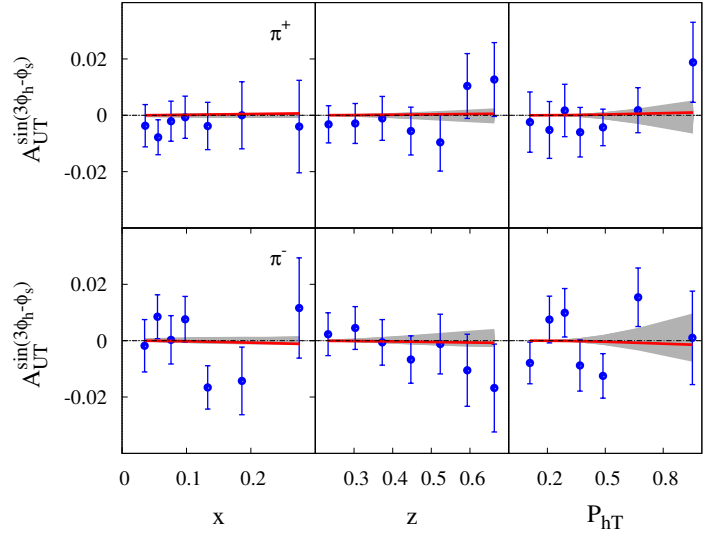}  
\vskip 1.cm
\caption{Preliminary HERMES \cite{Diefenthaler:2010zz,Schnell:2010zza,Pappalardo:2010zz} results on $A_{UT}^{\sin(3\phi_h-\phi_S)}$  as a function of $x,z,P_{hT}$ for $\pi^+$ and $\pi^-$. The  solid line corresponds to
the best fit and the shadowed region corresponds to the error corridor.}
\label{fig:aut}
\end{figure*}

The results of the description of preliminary experimental HERMES \cite{Diefenthaler:2010zz,Schnell:2010zza,Pappalardo:2010zz} data   
for $\pi^+$ and $\pi^-$ production on a proton target are presented in Fig.~\ref{fig:aut}.  Note that schematically for $\pi^+$ production on the proton target $F_{UT}^{\sin(3\phi_h-\phi_S)} \propto
 4 h_{1T}^{\perp u} H_{1}^{\perp fav} + h_{1T}^{\perp d} H_{1}^{\perp unfav}$ and because our result indicates that $h_{1T}^{\perp u} H_{1}^{\perp fav}  >0$ and $h_{1T}^{\perp d} H_{1}^{\perp unfav} >0$, the asymmetry is effectively enhanced and positive for $\pi^+$. Similarly for $\pi^-$ we have
 $F_{UT}^{\sin(3\phi_h-\phi_S)} \propto
 4 h_{1T}^{\perp (1) u} H_{1}^{\perp (1/2) unfav} + h_{1T}^{\perp (1) d} H_{1}^{\perp (1/2) fav} < 0$.
 
The smallness of the asymmetry in Fig.~\ref{fig:aut} is explained by suppression factor $z^2 P_{h T}^3$, because the
average values of HERMES are $\la z \ra \simeq 0.36$ and $\la P_{h T} \ra \simeq 0.4$ (GeV) and thus $z^2 P_{h T}^3 \simeq 0.008$ (GeV$^3$).
This makes possible values of the asymmetry be well below 1\%. \\

\begin{figure*}[htb]
\centering
\includegraphics[width=6cm,angle=-90]{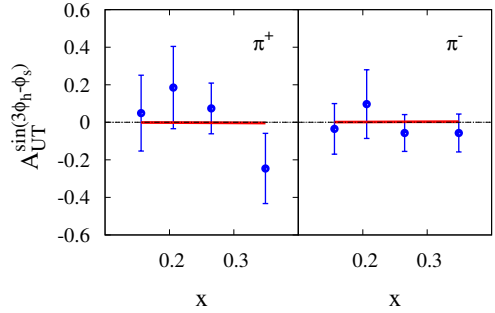} 
\vskip 0.5cm     
\caption{$A_{UT}^{\sin(3\phi_h-\phi_S)}$  measured by JLab \cite{Zhang:2013dow} on neutron ($^3$He) target as a function of $x$ for $\pi^+$ (left panel) and $\pi^-$ (right panel). The solid line corresponds to
the best fit and the shadowed region corresponds to the error corridor.}
\label{fig:autJLab}
\end{figure*}

Fit of the neutron data on $\pi^\pm$ production from JLab 6 \cite{Zhang:2013dow} is shown in Fig.~\ref{fig:autJLab}. The sign of the asymmetry for $\pi^+$ is negative, as
on neutron $F_{UT}^{\sin(3\phi_h-\phi_S)} \propto
 4 h_{1T}^{\perp d} H_{1}^{\perp fav} + h_{1T}^{\perp u} H_{1}^{\perp unfav} < 0$, and positive for $\pi^-$, as $F_{UT}^{\sin(3\phi_h-\phi_S)} \propto
 4 h_{1T}^{\perp d} H_{1}^{\perp unfav} + h_{1T}^{\perp u} H_{1}^{\perp fav} > 0$. Due to kinematical suppression the resulting asymmetry is very small, the measured asymmetry has very big errors and is compatible with our fit.

\subsection{Predictions for Jefferson Lab 12 GeV\label{sectionIVA}}
\begin{figure*}[htb]
     \vskip 0.5cm
     \includegraphics[width=6cm,angle=-90]{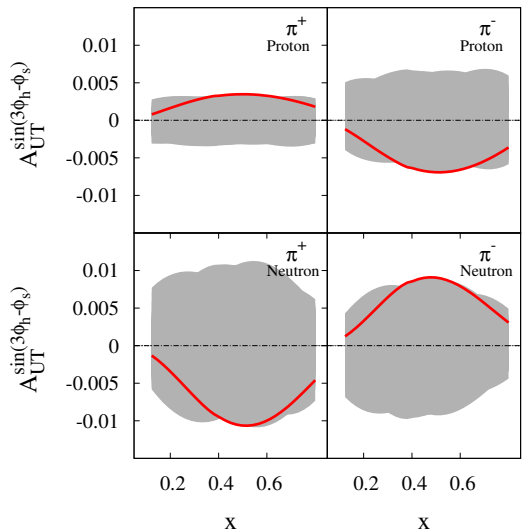} 
     \vskip 0.5cm
 \caption{Prediction of $A_{UT}^{\sin(3\phi_h-\phi_S)}$   at JLab 12 kinematics on proton (upper two plots) and neutron (bottom two plots)  targets as a function of $x$ for $\pi^+$ (left panels) and $\pi^-$ (right panels). The solid line corresponds to
the best fit and the shadowed region corresponds to the error corridor.}
\label{fig:JLab12}
\end{figure*}

\noindent
We present predictions for future measurements of $A_{UT}^{\sin(3\phi_h-\phi_S)}$  on a proton target at Jefferson Lab at 12 GeV in Fig.~\ref{fig:JLab12}.
We plot our prediction for $\pi^+$ production on a proton target assuming  $\langle z \rangle = 0.5$ and $\langle P_{hT} \rangle = 0.38$ GeV. We predict absolute value of the asymmetry 
of order of 1\%. Both positive and negative asymmetries are possible, current data prefer positive asymmetry for $\pi^+$ on the proton target (positive $u$-quark pretzelosity times positive favored Collins FF) and negative asymmetry for $\pi^-$ (positive $u$-quark pretzelosity times negative unfavored Collins FF). Signs of asymmetries on neutron target are reversed with respect to the proton target and absolute values are slightly higher.

\section{Comparison with other calculations\label{sectionV}} 
\noindent
Our results are of opposite sign if compared to model calculations of \cite{Kotzinian:2008fe,She:2009jq,Avakian:2008dz,Pasquini:2008ax,Pasquini:2009bv,Boffi:2009sh,Avakian:2010br}. Most models predict that $h_{1T}^{\perp u} < 0 $ and $h_{1T}^{\perp d} > 0$ while our best fit indicates that  $h_{1T}^{\perp u} > 0 $ and $h_{1T}^{\perp d} < 0$. However, as can be seen from Fig.~\ref{fig:pretzelosity}, our fit does not give a clear preference on the sign of pretzelosity.

The size of asymmetries is compatible with calculations of Ref.~\cite{She:2009jq}, where  asymmetries of order of 1\% for $\pi^+$ and 0.5\% for $\pi^-$  were found for JLab kinematics and can be compared to our findings in Fig.~\ref{fig:JLab12}.  Other calculations, for example, \cite{Kotzinian:2008fe} or \cite{Avakian:2008dz}, suggest bigger asymmetries up to 4\%-5\% for COMPASS kinematics and 2\%-5\% for JLab 12 kinematics. In contrast our calculations suggest that asymmetry at JLab 12 will be of order of 1\% at most. Future experimental measurements will be very important to clarify the sign and the size of pretzelosity and $A_{UT}^{\sin(3\phi_h-\phi_S)}$  asymmetry.

\section{Model relations and bounds for pretzelosity\label{sectionVI}}
\noindent
Positivity bound for the pretzelosity reads  \cite{Bacchetta:1999kz} 
\be
\left| h^{\perp (1)a}_{1T}(x) \right|   \le \frac{1}{2}(f_{1}^{a}(x) - g_{1}^{a}(x)) \, .
\label{eq:bound1}
\ee 
If the positivity  bound  is  combined   with  the Soffer bound $h_1^a(x) \le 1/2(f_{1}^{a}(x) + g_{1}^{a}(x))$ \cite{Soffer:1994ww} one obtains \cite{Avakian:2008dz}
\be
\left| h^{a}_{1}(x) \right|  + \left| h^{\perp (1)a}_{1T}(x) \right|   \le f_{1}^{a}(x)  \, .
\label{eq:bound2}
\ee 
In a certain class of models including bag models (see, e.g., \cite{Avakian:2008dz}) one obtains also the following model relations for the pretzelosity and transversity:
\ba
2 h^{a}_{1}(x)   = f^{a}_{1}(x) +    g_{1}^{a}(x) \;, \\
h^{\perp (1)a}_{1T}(x)   = h^{a}_{1}(x) -    f_{1}^{a}(x) \;.
\label{eq:model}
\ea 
or
\be
h^{\perp (1)a}_{1T}(x)   = g^{a}_{1}(x) -    h_{1}^{a}(x) \;.
\label{eq:model1}
\ee

\begin{figure*}[h]
  \begin{tabular}{c@{\hspace*{0.1cm}}c}
    \includegraphics[width=6cm]{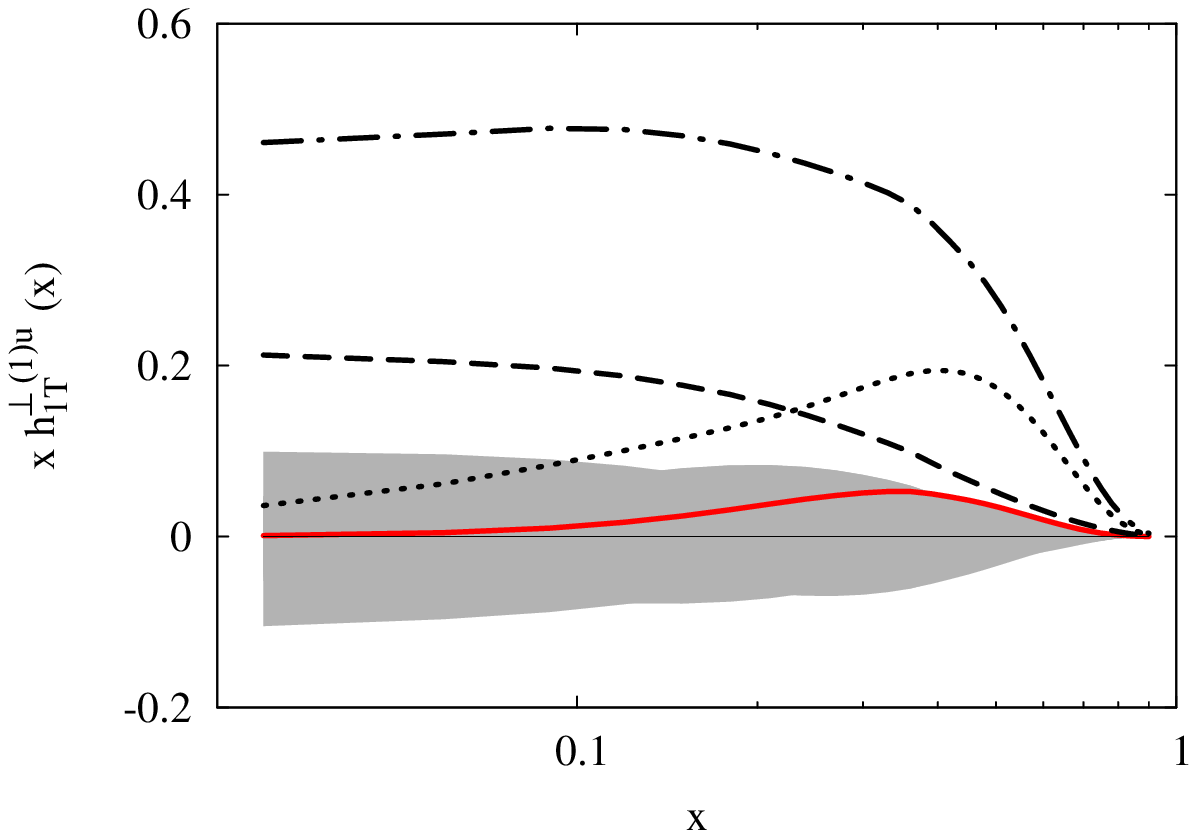}
    &
    \includegraphics[width=6cm]{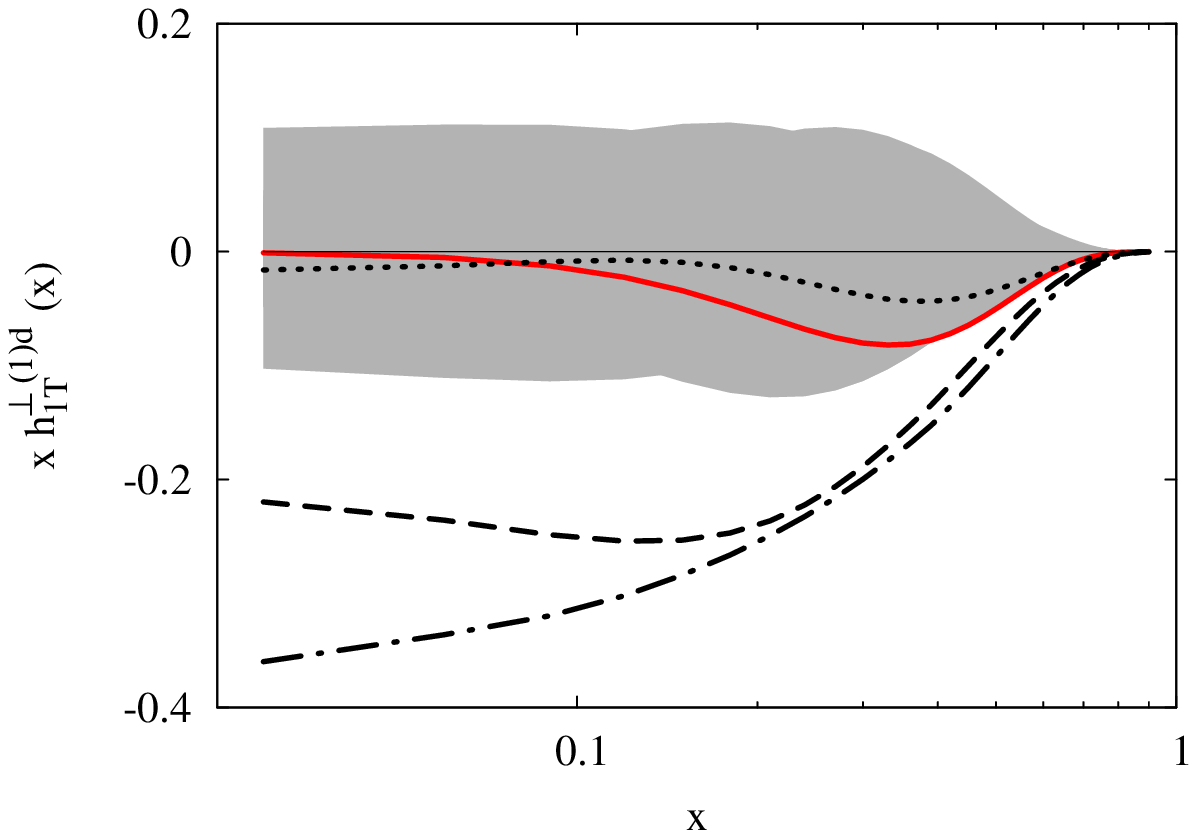}
  \\
  (a) & (b)
  \end{tabular}     
\caption{First moment of the pretzelosity distribution for up (a) and down (b) quarks. The red solid line corresponds to
the best fit and the shadowed region corresponds to the error corridor. The dotted line is the model relation Eq. \eqref{eq:model1} $h^{\perp (1)a}_{1T}(x)   = g^{a}_{1}(x) -    h_{1}^{a}(x)$, the thick dashed line is the positivity bound Eq. \eqref{eq:bound1} $\pm|\frac{1}{2}(f^{a}_{1}(x) -    g_{1}^{a}(x))|$,  and
the alternating dashed and the dotted line is the bound from Eq. \eqref{eq:bound2} $\pm|f_1^a(x) - \left| h^{a}_{1}(x) \right||$. Neither of the bounds is violated.}  \label{fig:pretzelosityandmodels}
\end{figure*}

Let us examine these model relations. Eq. \eqref{eq:model} implies that transversity saturates the Soffer bound \cite{Soffer:1994ww}. In fact we know that
phenomenological extraction of transversity is close to the bound, however the bound is not saturated (see, e.q., Ref.~\cite{Anselmino:2008sga}).
If the bound were saturated, then Eqs. (\ref{eq:model},\ref{eq:model1}) would simply read:
\be
h^{\perp (1)a}_{1T}(x)   = \frac{1}{2}(g^{a}_{1}(x) -    f_{1}^{a}(x)) \;;
\label{eq:model3}
\ee 
i.e., the positivity bound for the pretzelosity would be saturated as well.

In order to compare these model predictions with our results we plot in Fig.~\ref{fig:pretzelosityandmodels} the first moment of pretzelosity for up and down quarks and the 
results from Eq. \eqref{eq:model1} using transversity from  Ref.~\cite{Anselmino:2008sga} (the dotted line) and  positivity bound \eqref{eq:bound1} (the thick dashed  line).
One can see that  if one uses extracted  transversity  in Eq.  \eqref{eq:model1}, then the resulting pretzelosity violates the  positivity bound. We also plot $f_1^a(x) - \left| h^{a}_{1}(x) \right|$ (dot dashed lines). One can see that neither of positivity bounds Eqs.\eqref{eq:bound1}, \eqref{eq:bound2} is violated by 
our extracted pretzelosity.
The model relation of Eq.~\eqref{eq:model1} does violate one of the positivity bounds if transversity does not saturate Soffer bound. Numerical comparison 
of Eq.~\eqref{eq:model1} with extracted pretzelosity suggests that for up quarks there is a big discrepancy; in fact, our  parameterization is constructed to
satisfy the positivity bound while Eq.~\eqref{eq:model1} may violate it (compare Eq.~\eqref{eq:model3} that assumes saturation of bounds and model relation of Eq.~\eqref{eq:model1}). For down quarks the comparison is better, numerically results are similar, in this case the model relation of Eq.~\eqref{eq:model1} numerically satisfies the bound.
We also checked that if one fits the data without imposing positivity constraints when the extracted first moment of pretzelosity does not violate the positivity bound in the region of $x$ where experimental data are available, $0.0065 < x < 0.35$. At large values of $x$ violation is possible; however, this region is not constrained by the data.

\section{Quark Orbital Angular momentum\label{sectionVII}}
\noindent
Using the pretzelosity from the previous section, let us calculate quark OAM in the region of experimental data $0.0065 \le x \le 0.35$ 
\ba
{\cal L}_{z }^{a[x_{min},x_{max}]} = -\int_{x_{min}}^{x_{max}} d x \; h_{1T}^{\perp (1) a}(x, Q^2)\; .
\ea
Using the parameters with errors from Table~\ref{fitparI} we calculate the following values
at $Q^2 = 2.4$ GeV$^2$:
\ba
{\cal L}_{z }^{u [0.0065,0.35]} = -0.03^{+ 0.25}_{-0.10} \, ,\nonumber \\  
{\cal L}_{z}^{d [0.0065,0.35]}  = +0.05^{+ 0.49}_{-0.34}  \, . 
\ea
If we integrate over the whole kinematical region $0 < x < 1$ then we obtain
\ba
{\cal L}_{z}^{u [0,1]} = -0.06^{+ 0.38}_{-0.10} \, ,\nonumber \\  
{\cal L}_{z}^{d [0,1]} = +0.08^{+ 0.93}_{-0.60} \, .  
\ea
One notes that substantial value of the integral comes from unexplored high-$x$ and low-$x$ regions.

\section{Conclusions\label{conclusions}}
\noindent
We performed the first extraction of the pretzelosity distribution from preliminary COMPASS, HERMES,  and JLab experimental data. Even though
the present extraction has big errors we conclude that up-quark pretzelosity tends to be positive and down-quark pretzelosity
tends to be negative. This conclusion is not in agreement with models \cite{Kotzinian:2008fe,She:2009jq,Avakian:2008dz,Pasquini:2008ax,Pasquini:2009bv,Boffi:2009sh,Avakian:2010br} that predict negative up-quark pretzelosity and positive down-quark pretzelosity. We note that extracted pretzelosity has very big errors and allow for both positive and negative signs. Indeed, a vanishing asymmetry is very consistent with existing experimental data. Future experimental data from Jefferson Lab 12 \cite{Dudek:2012vr} will be essential for determination of the properties of the pretzelosity distribution.

The extracted pretzelosity can be related in a model dependent way to quark OAM and  at  $Q^2 = 2.4$ GeV$^2$ ${\cal L}_{z}^{u [0,1]} = -0.06^{+ 0.38}_{-0.10}$,
${\cal L}_{z}^{d [0,1]} = +0.08^{+ 0.93}_{-0.60} $.

\section*{Acknowledgments}
\noindent
We would like to thank Mauro Anselmino, Elena Boglione, Bakur Parsamyan, Gunar Schnell, Wally Melnitchouk, Alberto Accardi, and
 Pedro Jimenez-Delgado for help and fruitful discussions. We thank Gunar Schnell for discussion on experimental data and fitting procedures. We would like to thank the referee of this paper for his/her thoughtful comments that helped us to sharpen physics discussion.
 C.L. thanks the Department of Energy's Science Undergraduate Laboratory Internships (SULI) for support during his stay at Jefferson Lab.
This material is based upon work supported by the U.S. Department of Energy, 
Office of Science, Office of Nuclear Physics, under Contract No.~DE-AC05-06OR23177 (A.P.).


\end{document}